# Visualization use in qualitative research reports: Evolving media types and competing epistemologies

Visual displays in Qualitative Research

Jayrylle R. Jaylo

School of Interdisciplinary Arts & Sciences, University of Washington Bothell, jkrjx@uw.edu

MIA CHASTAIN

School of Interdisciplinary Arts & Sciences, University of Washington Bothell

ALLI NEMEC

School of Interdisciplinary Arts & Sciences, University of Washington Bothell, anemec04@gmail.com

CHRISTINA S. OUCH

School of Interdisciplinary Arts & Sciences, University of Washington Bothell

YARED ASEFA

School of Interdisciplinary Arts & Sciences, University of Washington Bothell, yared_asefa@icloud.com

MARCUS LI

School of Interdisciplinary Arts & Sciences, University of Washington Bothell, limar23@uw.edu

ANDREW UNG

School of Interdisciplinary Arts & Sciences, University of Washington Bothell, drewung@uw.edu

CALEB M. TRUJILLO[*]

School of Interdisciplinary Arts & Sciences, University of Washington Bothell, calebtru@uw.edu

Little is known about the representations used in qualitative research studies and why. A data-driven literature review was employed to explore the use of media in qualitative research reporting. A study by Verdinelli & Scagnoli (2013) was replicated and extended by conducting a content analysis of papers and figures published across three qualitative methods journals between 2020 and 2022. Figures were categorized by types (e.g., matrix-based, Venn diagrams, flowcharts) and documents were grouped by their epistemological stances (i.e., objectivist, subjectivist, or constructivist) before conducting a correspondence analysis and epistemic network analysis. Our findings suggest that (1) visual media have remained largely absent, (2) figure types have become more diverse and (3) the use of figure types is likely independent of epistemological stance but provide opportunities for further exploration. These findings provide a foundation for impactful integration of data visualization tools to enhance communicative power of findings across disciplines.

[*]Corresponding author: Caleb M. Trujillo, calebtru@uw.edu

# 1 INTRODUCTION

Data visualization continues to dominate research reports, presentations, and news media as a mode to communicate information, but little work has been done to understand appropriate representational considerations for qualitative research. In contrast to the cornucopia of graphics available for numeric values, the results of qualitative studies are often complex, nuanced, reflexive and abstract, making it difficult for researchers to translate themes, processes, and qualia into a visual media (Bauer & Sippel, 2024). These differences can be traced to their data sources and diverging traditions.

Qualia capture human experiences, culture, and meaning in the form of raw media recordings (e.g., audio, photos, video) while the latter, quanta, prioritizes measures, numbers and statistics (Terfloth et al., 2025; Point & Baruch, 2023). Thus, it comes as no surprise that visual media in qualitative reports are largely absent and definition of what makes an effective qualitative visual is lacking. About one in five articles in prominent qualitative research journals use visualizations (Verdinelli and Scagnoli, 2013; Henderson and Segal, 2013). Among these representations, matrices (i.e., tables), networks, flow charts, and conceptual forms tend to place the emphasis on quotes over counts. To better understand how visualizations have evolved, this research replicates and extends work from Verdinelli and Scagnoli's 2013 study of media. Our research in progress asks (1) how has the use of visual media changed in qualitative media and (2) can the variety of visualization use be explained by a studies use of epistemic basis. This study ultimately aims to identify where the gaps are between qualitative research and the use of visualizations, in hopes to create a bridge between the two.

## 1.1 Background

Historically, the "incommensurable" divide between qualitative and quantitative methodologies stem from the nature of their data and their community norms (Foster, 2023). Quantitative research attends to reliability, validity, confidence, significance, and error by reporting estimates of uncertainty. However, these values are not without challenges. The bias inherent in quantitative data has become a central issue in the propagation of data-driven models (D'Ignazio & Klein, 2020) and the rhetoric of data visualizations (Wolfe, 2010; Lisnic et al., 2024). However, the norms of qualitative traditions have developed ways to address these issues. Qualitative reports value trustworthiness, the plausibility of their interpretation, and the credibility of the research through activities such as detailed description and multivocality (Tracy & Hinrichs, 2017). Despite these challenges, qualitative visualizations continue to be integrated into qualitative and mixed-methods research due their ability to synthesize major theoretical concepts and reflexive practices (Morse, 2006; Verdinelli & Scagnoli, 2013; Guetterman et al., 2021; Bauer & Sippel, 2024).

Several intellectual bridges embrace an application and understanding of qualitative data visualization. Grounded theory, for instance, has proven productive in supporting visualization research (Diehl et al., 2022). It has played a role in developing data displays, such as a text-based analytic interface (Chandrasegaran et al., 2017), and in facilitating collaboration between critical GIS and feminist ethnography (Knigge and Cope, 2006). Visual joint displays, which integrate qualitative and quantitative research, have shown growth in mixed methods and methodological journals (Guetterman et al., 2021). Furthermore, textual information has been visualized in a variety of ways using lexical analysis, natural language processing (Nguyen et al., 2021, 2022; Liu et al., 2009; Wang et al., 2016), or statistical approaches such as those of quant ethnography (Shaffer, 2018; Abramson & Dohan, 2015). However, text-based visual analytics disregard shared community values for reliability and trustworthiness (Kucher et al., 2013). The disconnect between the norms and functions of qualitative data visualization provides an opportunity for deeper research.

## 1.2 Epistemological Stances

Among its unique functions, qualitative research has 'pillar' traditions that enhance or limit the appropriate and justifiable tenets of study designs. Operationally, tenets refer to the ideological beliefs embedded in the ontological, epistemological, and methodological properties of a research study (Guba & Lincoln, 2005; Moon & Blackman, 2014). In plain language, tenets are the nature of truth, knowledge, and process. For this study, we examine the role epistemology may play in shaping qualitative displays. Epistemology refers to how people create knowledge and what is possible to know. Its concern is how knowledge is produced or acquired and the extent of its applicability (Moon & Blackman, 2014). As epistemology frames one's research, it's important to consider how that affects the use of visualizations and its relationship with visualizations. For the purpose of this study, three main epistemological stances are considered: objectivism, constructivism, and subjectivism. Each team member coded a subset of articles for epistemologies by examining the content of the paper. Objectivism is the belief that knowledge is discoverable and learned externally, constructivism is the belief that knowledge is learned through social interactions, and subjectivism is the belief that knowledge is gained through individual ideas (Guba & Lincoln, 2005; Moon & Blackman, 2014). A careful examination of (1) the visualization practices of qualitative research methods and (2) their relation to underlying tenets will be paramount to designing effective qualitative data visualizations and understanding the qual-quant methodological divide.

## 2 METHODS

Similar to the original exploratory study, we identified articles published between 2020 and 2022 from the same three qualitative research journals: Qualitative Health Research, Qualitative Inquiry, and Qualitative Research. Following reporting norms (Moher et al., 2015), we collected a total of 1,256 papers and selected 10% of these papers. Documents were randomly selected with 40 papers per year for even distributions, for a total of 120 papers. Out of those papers, 55 papers had a minimum of one visual, giving us 152 visuals to analyze. We conducted a three-phase content analysis to analyze the visualizations and their current trends in qualitative research (Kuckart & Rädiker, 2023). For our initial pass, the original study's rubric was deductively applied to the sampled articles recording the total number of visualizations and the total number of tables. In order to facilitate consistent coding, we created a definition table with examples and written definitions for visualization types and epistemology categories (Appendix Table 1). For our second pass, we added multiple categories to address gaps in the original rubric: photographs, drawings, video, non-visual audio, and an other category—if the visualization didn't meet any of our definitions. Ambiguities were discussed and definitions were clarified. For our final pass, we continued to add data-based graphs, mixed media, and screenshots to our visualization types. To clarify a trend in photographs and drawings, which were not documented in the original study, we performed an ad hoc count with a random sample of 20 papers from the same journals across years 2007 to 2022, for a total of 320 articles.

Through our three-phase content analysis, we were able to create our final rubric. Once we created our final rubric, we estimated interrater reliability (IRR, O'Connor & Joffe, 2020) by randomly reassigning the team to a new set of 24 random papers from the 120 papers. We tested if two coders counted the same number of visualizations and their types to estimate IRR. Our results indicate that our rubric was highly reliable, with a 98.6% agreement. Interclass correlation revealed high agreement on counting figure types (ICC = 0.963) and moderate agreement on epistemology type (ICC = 0.633). Additionally, we categorize the research purpose of each visualization but chose to exclude these ratings from further analysis due to low ICC and IRR. To visualize these patterns, we used R to build our graphs and summary tables. Additionally, two exploratory analyses were performed using the same dataset: Epistemic Network Analysis (ENA, Shaffer, 2018) and Correspondence Analysis (CA, Husson et al., 2013). We applied ENA and CA to examine how these attributes co-occurred.

## 3 RESULTS

Our results explore trends in the types and epistemological roles of visualizations and how they reflect the changes qualitative research has gone through the past two decades.

### 3.1 Comparison of Trends

The comparison of visualization types from the original study to our study (Table 1) indicates a clear shift in the type of visualizations used. Similar to the original study, matrix displays continue to be the most dominant visualization. Although, the overall usage of them has gone down, from making up 60% of the visuals used in the original study, to only 33% of the visuals used in our study. Other visuals like flowcharts, networks, and modified Venn diagrams also declined, while box displays, ladders, and taxonomies became more common. For the past three years, photographs and matrix displays have become the most common. While the overall use has declined, the results suggest that the type of visualizations continue to diversify.

Table 1. Comparison table of the frequency of use for visualization types between the Verdinelli study and our study. The table calculates the percentage of each visualization type out of the total figures.

| Types of Display | 2007-2009 Frequency of Use | 2007-2009 Percentage | 2020-2022 Frequency of Use | 2020-2022 Percentage |
| --- | --- | --- | --- | --- |
| Matrix Display | 227 | 60% | 36 | 33% |
| Photographs | NA | NA | 14 | 13% |
| Box Display | 29 | 8% | 9 | 8% |
| Flowchart | 35 | 9% | 8 | 7% |
| Network | 48 | 13% | 7 | 6% |
| Ladder | 6 | 2% | 6 | 6% |
| Taxonomy | 8 | 2% | 6 | 6% |
| Data-Based Graphs | NA | NA | 5 | 5% |
| Metaphorical Visual Display | 4 | 1% | 5 | 5% |
| Drawings | NA | NA | 3 | 3% |
| Mixed Media | NA | NA | 3 | 3% |
| Modified Venn Diagram | 17 | 5% | 3 | 3% |
| Screenshots | NA | NA | 2 | 2% |
| Decision Tree Model | 3 | 1% | 1 | 1% |
| Non-Visual Audio | NA | NA | 1 | 1% |
| Total | 377 | 100% | 109 | 100% |

### 3.2 Visualization Usage

While the type of visualizations used are more diverse, the usage of visualizations in qualitative research remains sparse. Over half of the papers had no figures, while around 25 papers had 1-2 figures, and less than 15 papers had more than 3-5 figures (Fig. 1A). This shows that no figures were the most common finding in the papers. With a broader mix of visualization types, there's a steady, parallel increase in the use of both drawings and photographs in qualitative research papers from 2007 to 2022 (Fig. 1B). While there's an overall upward trend, there's also significant year-to-year fluctuations. Notably, there is a sharp spike in 2014, followed by a decline, and then another increase in 2022. While this trend isn't consistent, the overall growth suggests there is a growing openness with visual media in qualitative research, especially in formats like photographs and drawings.

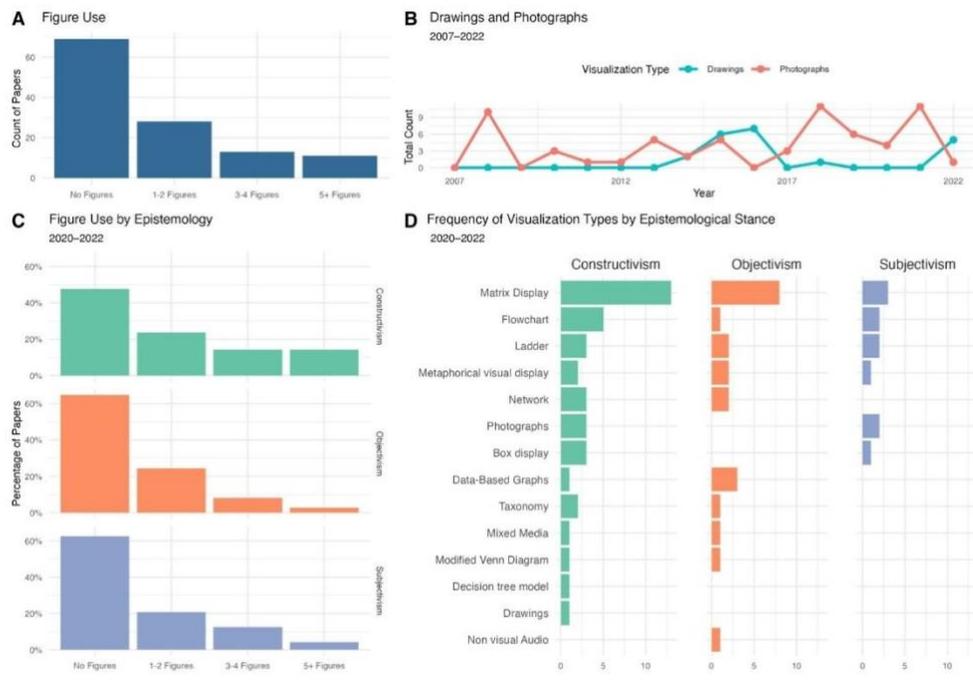

Figure 1. A. Count of papers by figure usage. This counts the number of papers within each figure category. It categorizes figure usage as no figures, 1-2 figures, 3-4 figures, and 5+ figures. B. Comparison of drawings and photographs over the years (2007-2022). The growth of photographs has been inconsistent, with peaks and declines over the years. The growth of drawings had a peak and is beginning to rise again. C. Percentage of papers by figure usage and epistemological stance. Constructivist papers show higher use of figures, whereas objectivism and subjectivism papers share similar patterns in frequency. D. Visualization types by epistemological stances (2020-2022). This figure counts the frequency of each visualization type in each epistemological stance.

### 3.3 Epistemologies

Although figure types varied independently with epistemological stances, we noted patterns worthy of description and future investigation (Fig. 1C, 1D). Across all three epistemology stances, matrix displays were the most common visual, again following the pattern of matrix usage. Along with matrix displays, metaphorical displays, ladders, and flowcharts were the only visualization types used across epistemological stances (Fig. 1D). Constructivist papers were the most common in the sample and utilized figures extensively. For example, more than 12% contained 5 or more figures. In contrast, over 60% of objectivist and subjectivist papers contained no figures. Meanwhile, objectivist papers used more data-based graphs but excluded photographs and drawings. Alternatively, subjectivism papers were limited and had the least number of figures as well.

Results from the ENA and CA support these descriptive statistics (Fig. 2, 3). The ENA revealed consistent patterns (Fig. 2), matrix displays associated centrally across epistemologies. Constructivist papers have the most connections, with varying degrees of strength to all of them. Meanwhile, results from CA suggested strong explanatory dimensions. The horizontal dimension strongly distinguished (73%) subjectivist and objectivist work, grouping photographs and no figure toward the subjectivist side and shifting modified Venn diagrams and Networks toward objectivists. Along the vertical dimension, constructivist studies are only somewhat distinguished, but media like drawings and decision trees strongly contributed and associated. However, statistical results suggest that the visualization types are likely independent from epistemology use. (Chi-squared: 11.63197, $p = 0.93$). This may be due in part to the limited sample of papers.

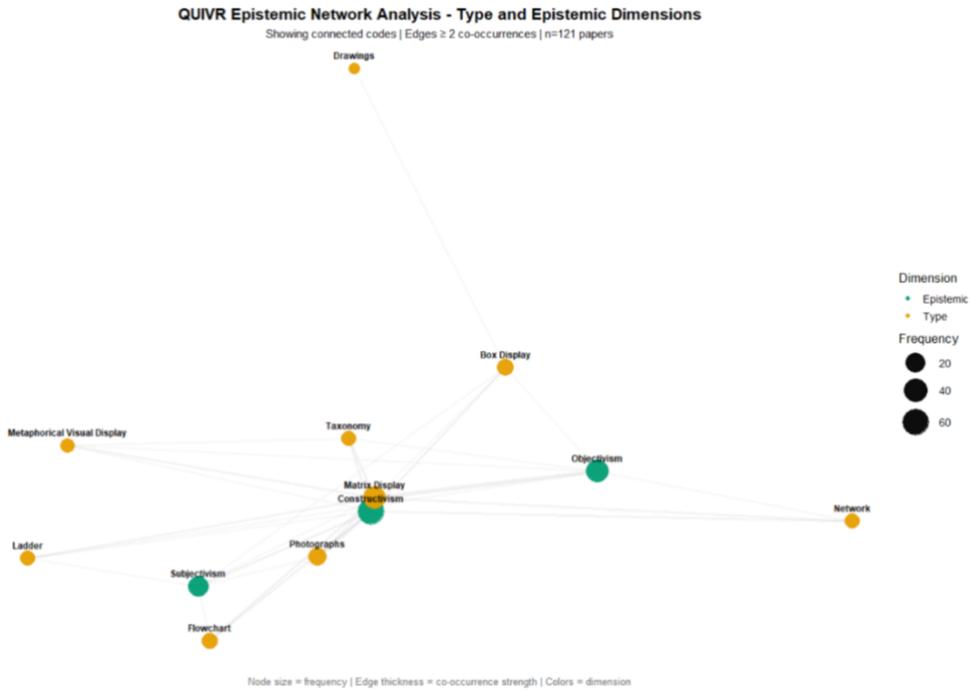

Figure 2. A network analysis on the relationship between epistemological stances and visualization type. The node size reflects the frequency of each variable, whereas the edge thickness shows the co-occurrence strength.

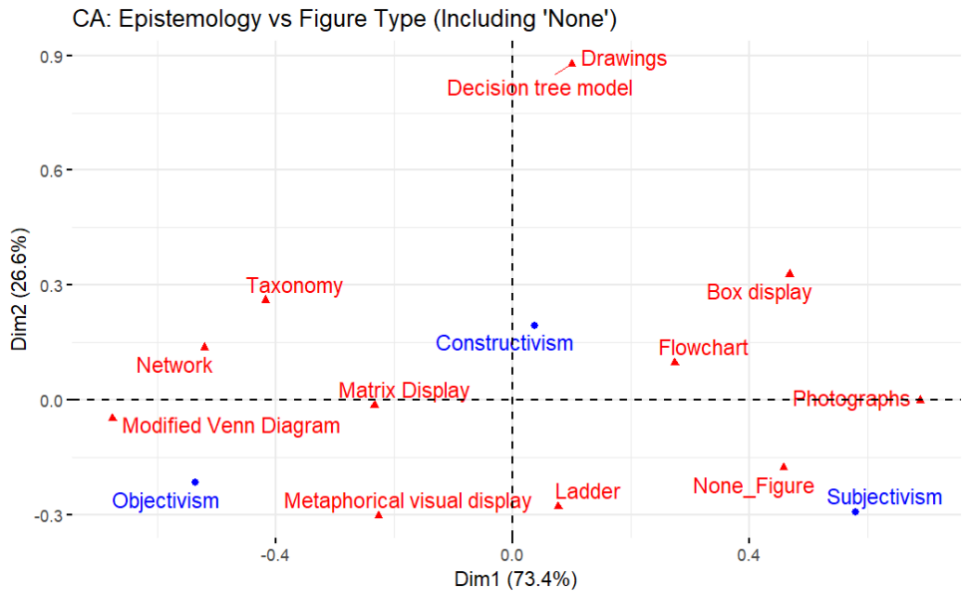

Figure 3. A correspondence analysis between an article's espoused epistemology and the visualization types present. Distance between points reflects the strength of association.

## 4 DISCUSSION

The ability to develop and interpret statistical graphics has been termed data visualization literacy—combining ideas from information literacy and visual literacy—and largely emphasizes quantitative media literacy (Börner et al., 2019; Ge et al., 2024a, 2024b; Hedayati et al., 2024). However, guidance for designers and budding qualitative researchers is lacking, which, if offered, could help speed communication and inform a new literacy regarding the usage of qualitative visualizations. We set out to understand the current usage and trends of visualizations in qualitative research. This preliminary research contributes findings on the evolving use of visualizations, an updated rubric to evaluate these graphics, and suggestions for future research.

Our findings reveal notable patterns. It further supports the relevant connection between epistemologies and visualization types. This also gives us insight into how qualitative researchers visualize complicated themes and findings in an organized, simplified manner. By analyzing visualizations and their uses, types and context, we are able to gain a deeper understanding of the current relationship between qualitative research and visual media. When comparing our results to the original study, there has been little change in the pervasiveness of visuals—about 20% of reports contain graphics—and matrix displays continue to be the most dominant visualization. This could be due to the nature of qualitative research; matrix displays provide a clean way to present and organize quotes, themes, and results. Understanding the role of matrix displays can give us insight into what is needed when translating qualitative work into visualizations. The deep reliance on text-based matrices suggests new opportunities for innovation due to similarity to quantitative approaches. While matrix displays are still common, box displays, ladders, and taxonomies are commonly seen in qualitative research.

The types of graphics authors are included are diversifying. The increased use of photographs and drawings are also a notable trend. The original study intentionally picked articles without photographs as it didn't match the definition they set for their visualizations. Our graph shows the potential if they had included photographs and drawings into their study. The declines and spikes of growth could be attributed to the accessibility of technology—smartphones grew in this period—and the overall acceptance of technology in research publications, a trend easily explored with a larger sample size. There could be a lack of expectations in the qualitative realm to utilize visualizations, leading researchers to have a difficult time in understanding the benefits of it and neglecting its potential. Visualizations give researchers the opportunity to represent how they think, what they value, and how they communicate.

Our modified rubric displayed strong results, giving us the opportunity to stay updated on the visualization trends. By following the visualization trends, it can help us to understand how to continue the normalization of visualizations in qualitative work. This rubric provides a potential tool and resource for researchers to better understand what visualizations can be helpful in communicating their ideas. This study is limited, with its constraint on the journal selection and year selection. This prevents us from understanding how visualizations are used in other qualitative disciplines and from seeing the trends between the original study and ours. A seamless integration between qualitative research and visualizations could provide us with new ways of understanding complex research topics. Moving forward, there is opportunity for stronger rubric definitions and criteria to develop stronger statistical analyses and comparisons. There is also an opportunity for diversifying the designs of visualizations, giving researchers the chance to connect their qualitative and quantitative ideas (Guetterman et al., 2021). While qualitative visualizing methods are not yet normalized, their type and use continue to evolve, indicating their growing role in qualitative research.

Future research should consider qualitative data visualization literacy as a theoretical construct. By combining qualitative considerations (e.g., theoretical frameworks, positionality, ontological views, epistemological stances, methodological approaches) and data visualization literacy (e.g., measurement type, scale, bias, interpretation, and construction), our future research seeks to develop a model of this practice by engaging disparate communities (Naeem et al., 2023) and use this model to recommend new methodologies (McMeekin et al., 2020). In doing so, we seek further calls to improve disciplinary diversity in visualization (Losev et al., 2022).

**ACKNOWLEDGMENTS**

The authors would like to express our gratitude to Kathleen Bowe, for stepping in as an additional advisor and providing us with new perspectives. The authors thank the University of Washington Bothell for the Scholarship, Research and Creative Practices Seed Grant to pursue this research and the UWB Student Academic Enhancement Fund, IAS Experiential Learning Gift Fund, and UW Undergraduate Research Conference Travel Award for travel support to attend CHI '26.

**REFERENCES**

[1] Abramson, C. M. and Dohan, D. 2015. Beyond text: Using arrays to represent and analyze ethnographic data. *Sociological Methodology* 45, 1 (2015), 272–319.

[2] Bauer, L. and Sippel, S. R. 2024. Illustrating qualitative research findings: The reflexive and epistemic potential of experimental visualization. *Geographica Helvetica* 79, 4 (2024), 373–389. https://doi.org/10.5194/gh-79-373-2024

[3] Börner, K., Bueckle, A., and Ginda, M. 2019. Data visualization literacy: Definitions, conceptual frameworks, exercises, and assessments. *Proceedings of the National Academy of Sciences* 116, 6 (2019), 1857–1864.

[4] Butler, D., Almond, J., Bergeron, R., Brodlie, K., and Haber, R. 1993. Visualization reference models. In *Proceedings of the IEEE Visualization Conference (VIS '93)*. IEEE, 337–342.

[5] Chandrasegaran, S., et al. 2017. Integrating visual analytics support for grounded theory practice in qualitative text analysis. *Computer Graphics Forum* 36, 3 (2017), 201–212.

[6] Davison, R. M., et al. 2024. The ethics of using generative AI for qualitative data analysis. *Information Systems Journal* (2024). https://doi.org/10.1111/isj.12504

[7] Diehl, A., Abdul-Rahman, A., Bach, B., El-Assady, M., Kraus, M., Laramee, R. S., and Chen, M. 2022. Characterizing grounded theory approaches in visualization. *arXiv preprint* arXiv:2203.01777.

[8] Foster, C. 2023. Methodological pragmatism in educational research: From qualitative–quantitative to exploratory–confirmatory distinctions. *International Journal of Research & Method in Education* 47, 1 (2023), 4–19. https://doi.org/10.1080/1743727X.2023.2210063

[9] Ge, L. W., Hedayati, M., Cui, Y., Ding, Y., Bonilla, K., Joshi, A., and Kay, M. 2024. Toward a more comprehensive understanding of visualization literacy. In *Extended Abstracts of the 2024 CHI Conference on Human Factors in Computing Systems (CHI '24)*. ACM, 1–7.

[10] Ge, L. W., Easterday, M., Kay, M., Dimara, E., Cheng, P., and Franconeri, S. L. 2024. V-FRAMER: Visualization framework for mitigating reasoning errors in public policy. In *Proceedings of the 2024 CHI Conference on Human Factors in Computing Systems (CHI '24)*. ACM, 1–15.

[11] Gilbert, J. K., Reiner, M., and Nakhleh, M., eds. 2010. *Visualization in Mathematics, Reading and Science Education*. Springer. https://doi.org/10.1007/978-90-481-8816-1

[12] Guetterman, T. C., Fàbregues, S., and Sakakibara, R. 2021. Visuals in joint displays to represent integration in mixed methods research: A methodological review. *Methods in Psychology* 5 (2021), Article 100080. https://doi.org/10.1016/j.metip.2021.100080

[13] Hedayati, M., Hunt, A., and Kay, M. 2024. From pixels to practices: Reconceptualizing visualization literacy. In *CHI 2024 Workshop: Toward a More Comprehensive Understanding of Visualization Literacy*.

[14] Henderson, S. and Segal, E. H. 2013. Visualizing qualitative data in evaluation research. *New Directions for Evaluation* 2013, 139 (2013), 53–71.

[15] Husson, F., Josse, J., Le, S., and Mazet, J. 2013. *FactoMineR: Multivariate Exploratory Data Analysis and Data Mining (Version 2.4)*.

[16] Knigge, L. and Cope, M. 2006. Grounded visualization: Integrating the analysis of qualitative and quantitative data through grounded theory and visualization. *Environment and Planning A* 38, 11 (2006), 2021–2037.

[17] Kucher, K., et al. 2013. An interdisciplinary perspective on evaluation and experimental design for visual text analytics: Position paper. In *Proceedings of the IEEE Workshop on Evaluation and Beyond – Methodological Approaches for Visualization (BELIV '13)*. IEEE, 28–35.

[18] Kuckartz, U. and Rädiker, S. 2023. *Qualitative Content Analysis*. SAGE Publications. https://doi.org/10.4135/9781036212940

[19] Lisnic, M., Lex, A., and Kogan, M. 2024. "Yeah, this graph doesn't show that": Analysis of online engagement with misleading data visualizations. In *Proceedings of the 2024 CHI Conference on Human Factors in Computing Systems (CHI '24)*. ACM, Article 199, 1–14. https://doi.org/10.1145/3613904.3642448

[20] Liu, S., Zhou, M. X., Pan, S., Qian, W., Cai, W., and Lian, X. 2009. Interactive, topic-based visual text summarization and analysis. In *Proceedings of the 18th ACM Conference on Information and Knowledge Management (CIKM '09)*. ACM, 543–552.

[21] Losev, T., et al. 2022. Embracing disciplinary diversity in visualization. *IEEE Computer Graphics and Applications* 42, 6 (2022), 64–71.

[22] McMeekin, N., Wu, O., Germeni, E., and Briggs, A. 2020. How methodological frameworks are being developed: Evidence from a scoping review. *BMC Medical Research Methodology* 20 (2020), Article 1.

[23] Moher, D., Shamseer, L., Clarke, M., Ghersi, D., Liberati, A., Petticrew, M., Shekelle, P., and Stewart, L. A. 2015. Preferred reporting items for systematic review and meta-analysis protocols (PRISMA-P) 2015 statement. *Systematic Reviews* 4, 1 (2015), Article 1. https://doi.org/10.1186/2046-4053-4-1

[24] Moon, K. and Blackman, D. 2014. A guide to understanding social science research for natural scientists. *Conservation Biology* 28, 5 (2014), 1167–1177. https://doi.org/10.1111/cobi.12326

[25] Morse, J. 2006. Diagramming qualitative theories. *Qualitative Health Research* 16, 9 (2006), 1163–1164.

[26] Nguyen, H. N., Trujillo, C. M., Wee, K., and Bowe, K. A. 2021. Interactive qualitative data visualization for educational assessment. In *Proceedings of the 12th International Conference on Advances in Information Technology*. ACM, 1–9.

[27] Nguyen, H. N., Dang, T., and Bowe, K. A. 2022. WordStream Maker: A lightweight end-to-end visualization platform for qualitative time-series data. *arXiv preprint* arXiv:2209.11856.

[28] O'Connor, C. and Joffe, H. 2020. Intercoder reliability in qualitative research: Debates and practical guidelines. *International Journal of Qualitative Methods* 19 (2020). https://doi.org/10.1177/1609406919899220

[29] Pandey, A. V., Manivannan, A., Nov, O., Satterthwaite, M., and Bertini, E. 2014. The persuasive power of data visualization. *IEEE Transactions on Visualization and Computer Graphics* 20, 12 (2014), 2211–2220. https://doi.org/10.1109/TVCG.2014.2346419

[30] Point, S. and Baruch, Y. 2023. (Re)thinking transcription strategies: Current challenges and future research directions. *Scandinavian Journal of Management* 39, 2 (2023), 101272.

[31] Shaffer, D. W. 2018. Epistemic network analysis: Understanding learning by using big data for thick description. In *The International Handbook of the Learning Sciences*. Routledge, 520–531.

[32] Terfloth, L., Lohmer, V., Kern, F., and Schulte, C. 2025. Transcription in computing education research: A review and recommendations. *Informatics in Education* 24, 2 (2025), 377–405. https://doi.org/10.15388/infedu.2025.09

[33] Tracy, S. J. and Hinrichs, M. M. 2017. Big tent criteria for qualitative quality. In *The International Encyclopedia of Communication Research Methods*. Wiley.

[34] Verdinelli, S. and Scagnoli, N. I. 2013. Data display in qualitative research. *International Journal of Qualitative Methods* 12, 1 (2013), 359–381. https://doi.org/10.1177/160940691301200117

[35] Wang, X., Liu, S., Chen, Y., Peng, T. Q., Su, J., Yang, J., and Guo, B. 2016. How ideas flow across multiple social groups. In *Proceedings of the IEEE Conference on Visual Analytics Science and Technology (VAST '16)*. IEEE, 51–60.

[36] Wolfe, J. 2010. Rhetorical numbers: A case for quantitative writing in the composition classroom. *College Composition and Communication* 61, 3 (2010), 452–475.

# A  APPENDIX

Table 1. Rubric used to categorize figures by visualization type and papers by epistemological stances. Categories were adapted, modified, and extended from Verdinelli and Scagnoli (2013).

| Code Category | Column Name | Definition |
| --- | --- | --- |
| Visualization type: Count of type per paper | Matrix Display | To cross two or more dimensions, variables, or concepts of relevance to the topic of interest |
|  | Network | To depict relationships between themes and subthemes or categories and subcategories |
|  | Flowchart | To illustrate directional flow and show pathways of different groups |
|  | Box display | To highlight a specific narrative considered important and frame it in a box |
|  | Modified Venn Diagram | To indicate shared or overlapping aspects of a concept, a category, or a process |
|  | Taxonomy | To classify or organize information |
|  | Ladder | To represent the dimensions of the progression of certain phenomenon through time or to show levels or stages |
|  | Metaphorical visual display | To depict in a metaphorical way the topics or themes found |
|  | Decision tree model | To describe options, decisions, and actions |
| | Screenshot | Capturing an image of data displayed on the screen of phones, computers, tablets, or other technology. An image of the actual screen interface. |
| | Data-based graphs | Visual representations such as bar graphs, networks, etc., using data |
| | Mixed Media | Mixed media includes collages, photographs with drawings layered, and other similar forms. A method that includes mixing different media forms (photographs, videos, graphs, etc.) into one figure |
| | Other | Visuals that are uncategorized |
| | Video | Recorded videos with visuals and audio |
| | Non-visual Audio | Recorded videos with no visuals, just audios, similar to podcasts |
| | Tables as figures | Tables that are labeled as a figure |
| | Photographs | An image of a person, place, or object captured by a camera |
| | Drawings | An illustration that is drawn by hand on paper or digitally. |
| Nature of knowledge: Assigned at the paper level | Objectivism epistemology | Knowledge is discoverable and is learned externally. Intention of discovery and measurement that requires us to leave our self/community to pursue an external truth/knowledge. Objects are the source of knowledge. |
| | Constructivism epistemology | Knowledge is learned through social connections, interactions, and exchanges. It's meant to emphasize learning through and with each other. Knowledge is developed through an active process within an individual or a group of individuals and this may nor may not accurately reflect the external objective world. Knowledge development is an internal or social process that does not require the knowledge seeker to prioritize the object. |
| | Subjectivism epistemology | Knowledge is gained through individual ideas and processes. It prioritizes your own individual experience and thoughts. Our knowledge is directly linked with our personal experiences and thoughts. |

.

## A.1 CRediT Roles

*Conceptualization - CT, JJ, MC, AN, YA, CO, ML, AU*
*Data Curation - CT, JJ, MC, AN, YA, CO, ML, AU*
*Formal Analysis - CT, JJ, MC, AN, YA, CO, ML, AU*
*Funding Acquisition - CT*
*Investigation - CT, JJ, MC, AN, YA, CO, ML, AU*
*Project Administration - CT, JJ*
*Resources - CT*
*Supervision - CT*
*Visualization - CT, JJ, MC, AN, YA, CO, ML, AU*
*Writing (Original Draft) - JJ*
*Writing (Review & Editing) - CT, JJ*